\documentclass[12pt]{article}
\usepackage{a4}
\usepackage[dvips]{graphicx}
\usepackage{amssymb, amsthm}

\newtheorem{theorem}{Theorem}[section]

\newtheorem{dfntn}{Definition}[section]

\def\RR{\mathbb{R}}

\begin{document}

\begin{center}
\Large \bf  Optimal systems of subalgebras for a nonlinear Black-Scholes equation
\end{center}
\begin{center}{Maxim Bobrov \\
\it Halmstad University, Box 823, 301 18 Halmstad, Sweden}\\[5pt]
\end{center}
\begin{abstract}
The main object of our study is a four dimensional Lie algebra
which describes the symmetry properties of a nonlinear
Black-Scholes model. This model implements a feedback effect which
is typical for an illiquid market. The structure of the Lie
algebra depends on one parameter, i.e. we have to do with a one-parametric family of algebras. We provide a classification of
these algebras using Patera--Winternitz method. Optimal systems of
one-, two- and three- dimensional subalgebras are described for
the family of symmetry algebras of the nonlinear Black-Scholes
equation. The optimal systems give us the possibility to describe a
complete set of invariant solutions to the equation.
\end{abstract}
[5pt]

\section{Introduction}
In \cite{bib:frey-patie-02} Frey and Patie  examined the feedback
effect of the option replication strategy of the large trader on
the asset price process. They obtained a new model by the introduction
of a liquidity coefficient  which depends on the current stock
price. The feedback-effect described  leads to a
 nonlinear version of the Black-Scholes partial differential equation,
\begin{eqnarray} \label{intur}
u_t+\frac{\sigma^2 S^2}2\frac{u_{SS}}{(1-\rho S \lambda (S)
u_{SS})^2}=0,
\end{eqnarray}
with $S \in [0,\infty), ~~ t \in [ 0,T ]. $ As usual, $S$ denotes
the price of the underlying asset and $u(S,t)$ denotes the
hedge-cost of the claim with a payoff $h(S)$ which will be defined
later. The hedge-cost is different from the price of the
derivatives product in illiquid markets. In the sequel  $t$ is the
time variable, $\sigma$ defines the volatility of the underlying
asset. The Lie group analysis of the equation (\ref{intur}) was provided in \cite{Bordag:2005}. By using this method the Lie point symmetries, the Lie symmetry algebras and groups to the corresponding
equations were found; see for details \cite{Bordag:2005} and
\cite{Lie, Olver, Ibragimov} for a general introduction to the
methodology.

\begin{theorem}[Bordag, \cite{Bordag:2005}]
The differential equation (\ref{intur}) with an arbitrary function
  $\lambda(S)$ possesses a trivial three dimensional Lie algebra $ Diff_{\Delta} (M)$ spanned by infinitesimal
  generators $$ v_1 =  \frac{\partial}{\partial t}, ~~
v_2 =  S\frac{\partial}{\partial u},~~ v_3 =
\frac{\partial}{\partial u}.$$ Only for the special form of the
function $\lambda(S) \equiv \omega S^k,$ where $ \omega, k\in {
R}$ equation (\ref{intur})
  admits a non-trivial four dimensional Lie algebra $L$ spanned by generators
\begin{eqnarray}\label{sysalg}
v_1 =  \frac{\partial}{\partial t}, ~~ v_2 =
S\frac{\partial}{\partial u},~~ v_3 = \frac{\partial}{\partial u},~~
v_4 = S \frac{\partial}{\partial S}+ (1-k) u
\frac{\partial}{\partial u} 
\end{eqnarray}
with commutator
relations
\begin{eqnarray}
~[v_1,v_2]=[v_1,v_3]=[v_1,v_4]=[v_2,v_3]=0,\nonumber\\
~[v_2,v_4]=-k v_2,~~~~[v_3,v_4]=(1-k) v_3. \label{alge}
\end{eqnarray}
\end{theorem}

In the paper \cite{YangZhangQu} authors try unsuccesses to construct an
optimal system of subalgebras for the symmetry algebra
(\ref{sysalg}). The authors used the method suggested in the
series of well known papers by P.Winternitz and Co \cite{PateraWinternitz, PateraWinternitzI}
where all three and four-dimensional Lie algebras where
classified.

 The investigation in \cite{YangZhangQu} contains some misprints and mistakes which
demand corrigendum. In that paper the structure of the optimal
system of subalgebras \cite{YangZhangQu} does not contain some one-dimensional
algebras. On the other sides the classification does not depend on the
parameter $k$ from which the structure of the algebra (\ref{sysalg}) deeply
depends. To be able to construct correct families of invariant
solutions we need a correct optimal system of subalgebras.\\ In
our paper we present the correct optimal system of one-, two-,
three- dimensional systems of subalgebras.

\section{Classification of the algebra $L$}
Let us consider the four-dimensional Lie algebra L (\ref{sysalg})
with the  commutator relations (\ref{alge}).
To classify this algebra we use the classification method which
was introduce by J. Patera and P. Winternitz in
\cite{PateraWinternitz}. Further we use the notations of this paper.\\
As we noticed before the structure of the algebra $L$ depends on the one real-valued parameter $k$. As it was remarked in \cite{Bordag:2005} the algebra posses a two-dimensional center by $k=0$ and $k=1$.\\
\paragraph{Case $k=0$.}
In this case the generators of the algebra $L$ take the form
\begin{equation}
v_1=\frac{\partial}{\partial t},~
v_2=x\frac{\partial}{\partial u},~
v_3=\frac{\partial}{\partial u},~
v_4=x\frac{\partial}{\partial x}+u\frac{\partial}{\partial u}.
\end{equation}
with the following commutator Table \ref{table1} \\
\begin{table}[h]
\begin{center}
\begin{tabular}{|c|c|c|c|c|}
\hline
{}&$v_1$&$v_2$&$v_3$&$v_4$\\
\hline
$v_1$&0&0&0&0\\
$v_2$&0&0&0&$0$\\
$v_3$&0&0&0&$v_3$\\
$v_4$&0&$0$&$-v_3$&0\\
\hline
\end{tabular}
\end{center}
\vspace{5pt}

\caption{The commutator table of the algebra $L$ in case $k=0$ }
\label{table1}
\end{table} \\
Let us introduce an algebra $L_{4.1}'$ with operators
\begin{equation}
e_1=-v_4,~e_2=v_3,~e_3=v_1,~e_4=v_2.
\end{equation}
Then $L_{4.1}'=A_2\oplus 2A_1$ with one nontrivial commutator relation  $[e_1,e_2]=e_2$.\\
\paragraph{Case $k=1$.}
This case leads to the algebra $L_{4.2}$ with generators
\begin{equation}
v_1=\frac{\partial}{\partial t},~
v_2=x\frac{\partial}{\partial u},~
v_3=\frac{\partial}{\partial u},~
v_4=x\frac{\partial}{\partial x}.
\end{equation}
which is isomorphic to the algebra $L_{4.2}'$ spanned on generators
\begin{equation}
e_1=v_4,~e_2=v_2,~e_3=v_1,~e_4=v_3.
\end{equation}
In this case algebra $L_{4.2}'$ has the decomposition $A_2\oplus 2A_1$ with one nontrivial commutator relations $[e_1,e_2]=e_2$.\\
\paragraph{Case $k>\frac12, k\neq 1$.}
In this case the algebra $L$ is isomorphic to the algebra $L_{4.3}'=A_{3.5}^{\alpha}\oplus A_1$ with the commutator relations $[e_1,e_3]=e_1,~~[e_2,e_3]=\alpha e_2$, where
\begin{equation}
e_1=v_2,~e_2=v_3,~e_3=-\frac1k v_4,~e_4=v_1 \label{k121}
\end{equation}
and $\alpha=\frac{k-1}{k}$.\\
\paragraph{Case $k<\frac12, k\neq 0$.}
This case leads us to the algebra $L$ which is isomorphic to $L_{4.4}=A_{3.5}^{\alpha}\oplus A_1$ with the commutator relations $[e_1,e_3]=e_1,~~[e_2,e_3]=\alpha e_2$, where
\begin{equation}
e_1=v_3,~e_2=v_2,~e_3=\frac{1}{1-k} v_4,~e_4=v_1 \label{k122}
\end{equation}
and $\alpha=\frac{k}{k-1}$ .\\
\paragraph{Case $k=\frac12$.}
In the last case the algebra  $L$ has generators of the following type
\begin{equation}
v_1=\frac{\partial}{\partial t},~
v_2=x\frac{\partial}{\partial u},~
v_3=\frac{\partial}{\partial u},~
v_4=x\frac{\partial}{\partial x}+\frac12 u\frac{\partial}{\partial u},
\end{equation}
and the commutator Table \ref{table3}\\
\begin{table}[h]
\begin{center}
\begin{tabular}{|c|c|c|c|c|}
\hline
{}&$v_1$&$v_2$&$v_3$&$v_4$\\
\hline
$v_1$&0&0&0&0\\
$v_2$&0&0&0&$-\frac12 v_2$\\
$v_3$&0&0&0&$\frac12 v_3$\\
$v_4$&0&$ \frac12 v_2$&$-\frac12 v_3$&0\\
\hline
\end{tabular}
\end{center}
\vspace{5pt}
\caption{The commutator table of the algebra $L$ in case $k=\frac12$ }
\label{table3}
\end{table}\\
and is isomorphic to the algebra $L_{4.5}'=  A_{3.4} \oplus A_1$ where generators are denoted by
\begin{equation}
e_1=v_3,~e_2=v_2,~e_3=2 v_4,~e_4=v_1.
\end{equation}
The nontrivial commutator relations are $[e_1,e_3]=e_1,~~[e_2,e_3]=-e_2$.
\section{An optimal system of subalgebras}
The main goal of this paper is to find a correct optimal system
of subalgebras for the Lie algebra $L$ (\ref{sysalg}). The procedure was
described by Pattera \& Winternitz in \cite{PateraWinternitz}. In
the paper all three and four-dimensional algebras were classified
and the optimal systems for these algebras were listed. We repeat
this algorithm for the algebra $L$ first in general case where $k\neq
0,\frac12,1$. In those two cases (\ref{k121}), (\ref{k122}) the Lie algebra $L$ is isomorphic to
the algebra $A_{3.5}^{\alpha}\oplus A_1$ with following commutator
Table \ref{gencomtable}
\begin{table}[h]
\begin{center}
\begin{tabular}{|c|c|c|c|c|}
\hline
{}&$e_1$&$e_2$&$e_3$&$e_4$\\
\hline
$e_1$&0&0&$e_1$&0\\
$e_2$&0&0&$\alpha e_2$&$0$\\
$e_3$&$-e_1$&$-\alpha e_2$&0&$0$\\
$e_4$&0&$0$&$0$&0\\
\hline
\end{tabular}
\end{center}
\vspace{5pt}
\caption{The commutator table of the algebra $L$ where $0<|\alpha|<1$.}
\label{gencomtable}
\end{table}\\
In the general case $L$ has one central
element and can be represent as a direct sum of one- and three-
dimensional Lie algebras
\begin{equation} \label{dirsum}
L=\{e_4\}\oplus L_{3},
\end{equation}
where $e_4$ is the central element of the algebra $L$ and $L_{3}= L\setminus \{e_4\}$.\\
The representation (\ref{dirsum}) simplify the procedure of
construction of an optimal system of subalgebras. \\We start with
construction of the corresponding system of subalgebras for both algebras
in \ref{dirsum} and then complete the study with the
investigation of non-splitting extensions. We follow the paper
\cite{PateraWinternitz} and describe a solution of this problem in a
step-by-step method introduced by the authors.\\
Step 1.
We find all subalgebras of $\{e_4\}$, it means we have $\{0\}$ and $\{e_4\}$.\\
Step 2. We have to classify all subalgebras of $L_3$ (\ref{dirsum}) under conjugation which is defined by an interior isomorphism of this algebra. This isomorphism can be presented by the adjoint representation.
{\dfntn[Olver, \cite{Olver}] Let $G$ be a Lie group. For each $g
\in G$, group conjugation $K_g(h) = ghg^{-1},~ h\in G$, determines
a diffeomorphism on $G$. Moreover, $K_g \circ K_{g'} = K_{gg'},
K_e = 1_G$, so $K_g$ determines a global group action of $G$ on
itself, with each conjugacy map $K_g$ being a group homomorphism:
$K_g(hh') = K_g(h)K_g(h')$ etc. The differential $dK_g:
TG|_h\rightarrow *TG|_{K_g(h)}$ is readily seen to preserve the
right-in variance of vector fields, and hence determines a linear
map on the Lie algebra of $G$, called the adjoint representation:
\begin{equation}
Ad~ g(v) = d K_g(v)
\end{equation}
}
The simplest way to find the adjoint representation is the Lie series
\begin{equation}
Ad(\exp(\varepsilon v))w=w-\varepsilon [v,w]+\frac{\varepsilon^2}{2!}[v,[v,w]]-\ldots \label{Ad}
\end{equation}
The adjoint representation table for the algebra $L$ (\ref{dirsum}) is rather simple and has a form given in Table \ref{adtable}
\begin{table}[h]
\begin{center}
\begin{tabular}{|c|c|c|c|c|}
\hline
{Ad}&$e_1$&$e_2$&$e_3$&$e_4$\\
\hline
$e_1$&$e_1$&$e_2$&$e_3-\varepsilon e_1$&$e_4$\\
$e_2$&$e_1$&$e_2$&$e_3-\alpha \varepsilon e_2$&$e_4$\\
$e_3$&$e^{\varepsilon} e_1$&$e^{\alpha \varepsilon} e_2$&$e_3$&$e_4$\\
$e_4$&$e_1$&$e_2$&$e_3$&$e_4$\\
\hline
\end{tabular}
\end{center}
\vspace{5pt}
\caption{The adjoint representation table of the algebra $L=A_{3.5}^{\alpha}\oplus A_1$ with $(i,j)$-th entry indicate $Ad(\exp(\varepsilon e_i))e_j$ element.
}
\label{adtable}
\end{table} \\
By using the adjoint representation (\ref{Ad}) we classify all subalgebras of $L_3$ (\ref{dirsum}) under conjugacy.
\paragraph{One-dimensional subalgebras.}
Firstly we consider one-dimensional subalgebras of the general type $$A=\{a e_1+b e_2+c e_3\},$$
where $a,~b,~c$ are arbitrary constants.
If $c\neq 0$ then we can use the first and the second lines of the Table \ref{adtable}. We obtain
\begin{equation}
Ad(\exp(\xi e_1+\zeta e_2))A= (a-c\xi) e_1 + (b-c\alpha \zeta) e_2+ c e_3, \label{genA}
\end{equation}
choosing $\xi=\frac ac,~~\zeta=\frac {b}{c\alpha }$ we prove that $A$ is isomorphic to $e_3$. \\
If $c=0$ in (\ref{genA}) we have three cases to study. \\If $a\neq 0,~b=0$ then $A$ is isomorphic to $e_1$. If $a=0,~b\neq0$ then $A$ is isomorphic to $e_2$. The last case we obtain if $ab\neq 0$ then we can use the third line of the Table \ref{adtable} and obtain after an action of the adjoint representation
\begin{equation}
Ad(\exp(\xi e_3))A= a e^{\xi} e_1 + be^{\alpha\xi} e_2.
\end{equation}
Using the scaling on $a e^{\xi}$ and choosing $\xi=\frac {1}{\alpha-1}
\log \left|\frac {a}{b} \right|$ we prove that $A$ is isomorphic to the algebra generated by
$\{e_1\pm e_2\}$.

Collecting all previous results we obtain that the optimal system of one-dimensional subalgebras of $L_3$ contains following subalgebras
\begin {equation}
\{0\},~~\{e_1\},~~ \{e_2\},~~ \{e_3\},~~\{e_1 \pm e_2\}.\label{onedim}
\end{equation}
\paragraph{Two-dimensional subalgebras.}
Let us consider now two-dimensional subalgebras of $L_{3}$. Let $B$ be one of the one-dimensional subalgebras (\ref{onedim}) and $A=\{a e_1+ b e_2+c e_3\}$. For a subalgebra $M=B+A$ we demand that $[A,B]\subset M$.\\
Let $B=\{e_1\}$ then without loss of generality we can represent $A$ in the form $\{a e_2+ b e_3\}$.  Let $b\neq 0$, by using of the second line of the adjoint representation Table \ref{adtable} we prove that $A$ is isomorphic to $\{e_3\}$. If $b=0$ then $A=\{e_2\}$. In this case we obtain two subalgebras
\begin{equation}
\{e_1,e_2\},~\{e_1, e_3\},
\end{equation}
which are non-isomorphic to each other.
In the same way we obtain subalgebras $\{e_1,e_2\},~\{e_2, e_3\}$ in case $B=e_2$.\\
Let $B=\{e_3\}$ then without loss of generality we can represent $A=\{a
e_1+b e_2\}$. Let us check the commutator relations
$$
[ a e_1+b e_2,e_3]=a e_1+\alpha b e_2.
$$
We see that $A+B$ is an algebra just under condition $ab=0$. On this way we obtain the two subalgebras
\begin{equation}
\{e_1,e_3\},~\{e_2, e_3\}
\end{equation}
In case $B=\{e_1\pm e_2\}$ we choose $A=\{e_3\}$. Then $[ e_1\pm e_2,e_3]=e_1\pm\alpha e_2$ is not an algebra.

%
The optimal system of the one- and two-dimensional subalgebras of $L_3$ contains following subalgebras
\begin{equation} 
\{0\},~~\{e_1\},~~ \{e_2\},~~ \{e_3\},~~\{e_1 \pm e_2\},~~\{e_1,e_2\},~~\{e_1,e_3\},~~\{e_2,e_3\}.\label{optsys}
\end{equation}
Step 3. We have to find all splitting extensions of the algebra $\{e_4\}$. To do this we have to find all subalgebras $N_a$ of $L_3$ such that
\begin{equation}
[e_4,~N_a]\subseteq N_a
\end{equation}
and classify all such subalgebras under $Nor_L e_4$.
{\dfntn[Ovsyannikov, \cite{ovsiannikov}] Let $N$ be a subalgebra of the Lie algebra $L$. By the normalizer $Nor_L N$ of $N$ in $L$ we mean the maximal subalgebra of $L$ containing $N$ in which $N$ is an ideal, i.e.
\begin{equation}
Nor_L N=\{y\in L: [y,x]\in N~~ \forall x\in N\}.
\end{equation}
}

As soon as $e_4$ is a central element
and $Nor_L e_4=L$ any adjoint representation does not affect on $e_4$ and $N_a$ is any subalgebra of $L_3$. This step is trivial and we obtain the subalgebras of the type $\{e_4,S\}$ where $S$ running through all subalgebras (\ref{optsys}).\\

Step 4. We have to find all subalgebras of type
\begin{equation}
\left\{e_4+\sum\limits_{i} a_i e_i, N_a \right\},
\end{equation}
where $N_a$ is a subalgebra of $L_3$ such that $Nor_L N_a$ is not contained in $L_3$, $a_i\in\RR$ are not all equal to zero and the generators $e_4+\sum\limits_{i} a_i e_i$ are not conjugate to $e_4$. Since $e_4$ is a central element of $L$ all of those conditions are satisfied.
Let $N_a$ running through the list of algebras (17) and let $A=e_4+\sum\limits_{i} a_i e_i$.
Let first $N_a=\{0\}$. This case is trivial because $e_4$ is the central element and the procedure was described on the Step 2. We obtain four subalgebras
\begin{equation}
\{a e_1+ e_4\},~~ \{a e_2+ e_4\},~~ \{a e_3+ e_4\},~~\{a(e_1 \pm e_2)+e_4\}, ~~a\neq 0
\label{bla}
\end{equation}
If we scale by $\frac1a$ all of these subalgebras (\ref{bla}) we see that for the two first subalgebras we can use  adjoint representation generated by $e_1$ and $e_2$ to reduce these subalgebras to simplest one. We obtain from Table \ref{adtable} following two subalgebras
\begin{equation}
\{e_1+ b e^{-\varepsilon} e_4\},~~ \{ e_2+ b e^{-\alpha \varepsilon} e_4\},
\end{equation}
where $b=\frac1a\neq 0$.
Choosing $\varepsilon = \log{|\frac1b|}$ in the first case and $\varepsilon = \frac{1}{\alpha}\log{|\frac1b|}$ in the second one, we finally obtain the following list of one dimensional non-splitting extensions
\begin{equation}
\{e_1\pm e_4\},~~ \{ e_2\pm e_4\},~~ \{ e_3+ a e_4\},~~\{e_1 \pm e_2+a e_4\},~~a\neq 0.
\end{equation}
Let us now consider two-dimensional non-splitting extensions. To simplify these procedures we use as $N_a$ the subalgebras of the list (\ref{optsys}). We notice that under action of the adjoint representation the general form of $e_4+\sum\limits_{i} a_i e_i$ is hold.

 Let $N_a$ be equal to $\{e_1\}$ then without loss of generality we can represent $A$ as $\{e_4+a_2 e_2 +a_3 e_3\}$. If $a_3$ is not equal to zero we can use the second line of the adjoint representation Table \ref{adtable} and reduce the algebra $A$ to $e_4+a e_3$. We obtain the following subalgebra

\begin{equation}
\{e_3+a e_4, e_1\}, a\neq 0.
\end{equation}
In the case $a_3=0$ we rewrite $A=\{a e_4+e_2\}$ and use the third line of the adjoint representation Table \ref{adtable} to obtain
\begin{equation}
\{e^{\alpha \varepsilon}e_2+a e_4, e^{\varepsilon}e_1\}
\end{equation}
or
\begin{equation}
\{e_2+ae^{-\alpha \varepsilon} e_4, e^{(1-\alpha)\varepsilon}e_1\}, a\neq 0.
\end{equation}
 By choosing $\varepsilon=\frac{1}{\alpha}\log{|a|}$ and scaling the second generator of the algebra above by the corresponding constant we obtain the following algebra
\begin{equation}
\{e_2\pm e_4, e_1\}.
\end{equation}
The same procedure in the case $N_a=e_2$ lead us to the non-isomorphic subalgebras
\begin{equation}
\{e_1\pm e_4, e_2\},~~\{e_3+a e_4, e_2\}, a\neq 0.
\end{equation}
Let us consider the case $N_a=\{e_3\}$, then we can
choose $A=\{e_4+a_1 e_1+a_2 e_2\}$. Note that $A+ N_a$ is an algebra
just in case $a_1 a_2=0$. Those subalgebras were considered in the
previous cases. Let $N_a=\{e_1 \pm e_2\}$ then $\{N_a, e_4+a_1
e_1+a_2 e_2+a_3 e_3\}$ is an algebra only if $a_3=0$.
Without loss of generality we can represent $A$ as an algebra generated by $\{a e_4+e_1\}$ then by
using the third line of the Table \ref{adtable} we see that algebras $\{A, N_a\}$ are isomorphic
to the following algebra
\begin{equation}
\{e_1\pm e_4, e_1+ a e_2\},
\end{equation}
where $a\in\RR$.
Note that the case $a=0$ we consider on the third step. Finally we obtain the following subalgebra
\begin{equation}
\{e_1\pm e_4, a e_1+ e_2\},
\end{equation}
where $a\neq 0$.\\
Now we consider case $N_a=\{e_1,e_2\}$ here we can represent $A$ as $\{a e_4+e_3\}$
and obtain the following three dimensional subalgebra
\begin{equation}
\{e_3+a e_4,  e_1, e_2\},
\end{equation}
where $a\neq0$. It is easy to see that the other choices of $N_a$ do not provide any other non similar subalgebras.\\

We obtain the following list of the optimal system of
subalgebras to the algebra $L$ (\ref{sysalg})
\begin{eqnarray*}
&{}&\{0\},~~\{e_1\},~~ \{e_2\},~~ \{e_3\},~~\{e_1 \pm e_2\},~~\{e_1,e_2\},~~\{e_1,e_3\},~~\{e_2,e_3\}, \\ &{}&\{e_4\},~~\{e_1,e_4\},~~ \{e_2,e_4\},~~ \{e_3,e_4\},~~\{e_1 \pm e_2,e_4\},~~\{e_1,e_2,e_4\},\\ &{}&\{e_1,e_3,e_4\},~~\{e_2,e_3,e_4\},~~\{e_1\pm e_4\},~~ \{ e_2\pm e_4\},~~ \{ e_3+ a e_4\},\\ &{}&\{e_1 \pm e_2+a e_4\},~~\{e_3+a e_4, e_1\},~~\{e_2\pm e_4, e_1\},~~\{e_1\pm e_4, e_2\}, \\ &{}&\{e_3+a e_4, e_2\}, \{e_1\pm e_4, a e_1+ e_2\}, ~~\{e_3+a e_4,  e_1, e_2\}.
\end{eqnarray*}
Finally we obtain the complete optimal system of subalgebras of Lie algebra $L$ (see Table \ref{opsys}).

\begin{table}[h]
\begin{center}
\begin{tabular}{|c|c|}
\hline
{Dimension}&{Subalgebras}\\
\hline
$1$&$\{e_2\},~~ \{e_4\},~~\{e_1 +a e_2\},~~\{e_1+\epsilon e_4\},$\\
&$\{e_2+\epsilon e_4\},~~ \{e_3+a e_4\},~~\{e_1 +\epsilon e_2+a e_4\}$\\
$2$&$\{e_1,e_2\},~~\{e_1,e_4\},~~\{e_2,e_4\},~~\{e_3,e_4\},~~\{e_1 +\epsilon e_2,e_4\}$\\
&$\{e_2+\epsilon e_4,e_1\},~~\{e_1+\epsilon e_4, a e_1+ e_2\},~~\{e_3+a e_4, e_1\},~~\{e_3+a e_4, e_2\}$\\
$3$&$\{e_1,e_2,e_4\},~~\{e_1,e_3,e_4\},~~\{e_2,e_3,e_4\},~~\{e_1,e_2,e_3+a e_4\},~~$\\
\hline
\end{tabular}
\end{center}
\vspace{5pt}
\caption{The optimal system of subalgebras of the algebra $L$ (\ref{sysalg}) in case $k\neq 0,\frac12 ,1$, were $a\in \RR,~~\epsilon=\pm1$. }
\label{opsys}
\end{table}
\newpage
We remark now that in  case $k=\frac12$ the structure of the algebra $L$ is the same as in the case above hence the optimal system of subalgebras is the same.\\
For $k=0$ or $k=1$  we obtain the following system of subalgebras by the similar procedure
\begin{table}[h]
\begin{center}
\begin{tabular}{|c|c|}
\hline
{Dimension}&{Subalgebras}\\
\hline
$1$&$\{e_2\},~~\{e_3 \cos {\varphi}+e_4\sin{\varphi}\},$\\
&$\{e_1+ a(e_3 \cos {\varphi}+e_4\sin{\varphi})\},$\\
&$\{e_2 +\epsilon(e_3 \cos{\varphi}+e_4\sin{\varphi})\}$\\
$2$&$\{e_1+a(e_3 \cos {\varphi}+e_4\sin{\varphi}),e_2\},~~\{e_3,e_4\},$\\
&$\{e_1 +a(e_3 \cos {\varphi}+e_4\sin{\varphi}),e_3 \sin{\varphi}-e_4\cos{\varphi}\},$\\
&$\{e_2 +\epsilon(e_3 \cos{\varphi}+e_4\sin{\varphi}),e_3 \sin{\varphi}-e_4\cos{\varphi}\},$\\
&$\{e_2,e_3 \sin {\varphi}-e_4\cos{\varphi}\}$\\
$3$&$\{e_1,e_3,e_4\},~~\{e_2,e_3,e_4\},$\\
&$\{e_1 +a(e_3 \cos {\varphi}+e_4\sin{\varphi}),e_3 \sin {\varphi}-e_4\cos{\varphi},e_2\},$\\
\hline
\end{tabular}
\end{center}
\vspace{5pt}
\caption{The optimal system of subalgebras of the algebra $L$ (\ref{sysalg}) in case $k= 0$ or $k=1$ where $a\in \RR,~~\epsilon=\pm1,~~0\leq\varphi\leq\Pi $. }
\label{opsys2}
\end{table}
\newpage
\section{Conclusion}
In this chapter we return to our original algebra $L$ (\ref{sysalg}) and introduce the optimal system of subalgebras Table \ref{opsys} and Table \ref{opsys2} in original generators.

\begin{table}[h]
\begin{center}
\begin{tabular}{|c|c|c|}
\hline
{Parameter}&{Generators}&{Optimal System}\\
&&{of subalgebras}\\
\hline
$k=0$&
$
{\scriptstyle
v_1=\frac{\partial}{\partial t},~
v_2=x\frac{\partial}{\partial u},~
}
$
&

${\scriptstyle
\{v_3\},~~\{v_1 \cos {\varphi}+v_2\sin{\varphi}\},~\{v_4+ a(v_1 \cos {\varphi}+v_2\sin{\varphi})\},}$\\
&

${\scriptstyle
v_3=\frac{\partial}{\partial u},~
v_4=x\frac{\partial}{\partial x}+u\frac{\partial}{\partial u}.
}
$
&
${\scriptstyle
\{v_3 +\epsilon(v_1 \cos{\varphi}+v_2\sin{\varphi})\},~\{v_1+a(v_1 \cos {\varphi}+v_2\sin{\varphi}),v_3\},
}$\\
&&
${\scriptstyle
\{v_1,v_2\},~\{v_4 +a(v_1 \cos {\varphi}+v_2\sin{\varphi}),v_1 \sin{\varphi}-v_2\cos{\varphi}\},
}$\\
&&
${\scriptstyle
\{v_3 +\epsilon(v_1 \cos{\varphi}+v_2\sin{\varphi}),v_1 \sin{\varphi}-v_2\cos{\varphi}\},~\{v_3,v_1 \sin {\varphi}-v_2\cos{\varphi}\},
}$\\
&&
${\scriptstyle
\{v_4,v_1,v_2\},~\{v_3,v_1,v_2\},~\{v_4 +a(v_1 \cos {\varphi}+v_2\sin{\varphi}),v_1 \sin {\varphi}-v_2\cos{\varphi},v_3\}.
}$\\
\hline
$k=1$&
$
{\scriptstyle
v_1=\frac{\partial}{\partial t},~
v_2=x\frac{\partial}{\partial u},~
}
$
&

${\scriptstyle
\{v_2\},~~\{v_1 \cos {\varphi}+v_3\sin{\varphi}\},~\{v_4+ a(v_1 \cos {\varphi}+v_3\sin{\varphi})\},}$\\
&

${\scriptstyle
v_3=\frac{\partial}{\partial u},~
v_4=x\frac{\partial}{\partial x}.
}
$
&
${\scriptstyle
\{v_2 +\epsilon(v_1 \cos{\varphi}+v_3\sin{\varphi})\},~\{v_1+a(v_1 \cos {\varphi}+v_3\sin{\varphi}),v_2\},
}$\\
&&
${\scriptstyle
\{v_1,v_3\},~\{v_4 +a(v_1 \cos {\varphi}+v_3\sin{\varphi}),v_1 \sin{\varphi}-v_3\cos{\varphi}\},
}$\\
&&
${\scriptstyle
\{v_2 +\epsilon(v_1 \cos{\varphi}+v_3\sin{\varphi}),v_1 \sin{\varphi}-v_3\cos{\varphi}\},~\{v_2,v_1 \sin {\varphi}-v_3\cos{\varphi}\},
}$\\
&&
${\scriptstyle
\{v_4,v_1,v_3\},~\{v_1,v_2,v_3\},~\{v_4 +a(v_1 \cos {\varphi}+v_3\sin{\varphi}),v_1 \sin {\varphi}-v_3\cos{\varphi},v_2\}.
}$\\
\hline
$k<\frac12,~k\neq 0$
&
${\scriptstyle
v_1=\frac{\partial}{\partial t},~
v_2=x\frac{\partial}{\partial u},~
v_3=\frac{\partial}{\partial u},
}
$
&
${\scriptstyle
\{v_2\},~ \{v_1\},~\{v_3 +a v_2\},~\{v_3+\epsilon v_1\},~\{v_2+\epsilon v_1\},~ \{v_4+a v_1\},
}
$\\
&
$
{\scriptstyle

v_4=x\frac{\partial}{\partial x}+(1-k)u\frac{\partial}{\partial u}.
}
$
&${\scriptstyle
\{v_3 +\epsilon v_2+a v_1\},~\{v_3,v_2\},~\{v_3,v_1\},~\{v_2,v_1\},~\{v_4,v_1\},~\{v_3 +\epsilon v_2,v_1\}
}
$
\\
&&
${\scriptstyle
\{v_2+\epsilon v_1,v_3\},~\{v_3+\epsilon v_1, a v_3+ v_2\},~\{v_4+a v_1, v_3\},~\{v_4+a v_1, v_2\},
}
$\\
&&
${\scriptstyle
\{v_3,v_2,v_1\},~\{v_3,v_4,v_1\},~\{v_2,v_4,v_1\},~\{v_3,v_2,v_4+a v_1\}.
}
$\\
\hline
$k=\frac12$
&
${\scriptstyle
v_1=\frac{\partial}{\partial t},~
v_2=x\frac{\partial}{\partial u},~
v_3=\frac{\partial}{\partial u},
}
$
&
${\scriptstyle
\{v_2\},~ \{v_1\},~\{v_3 +a v_2\},~\{v_3+\epsilon v_1\},~\{v_2+\epsilon v_1\},~ \{v_4+a v_1\},
}
$\\
&
$
{\scriptstyle

v_4=x\frac{\partial}{\partial x}+\frac12 u\frac{\partial}{\partial u}.
}
$
&${\scriptstyle
\{v_3 +\epsilon v_2+a v_1\},~\{v_3,v_2\},~\{v_3,v_1\},~\{v_2,v_1\},~\{v_4,v_1\},~\{v_3 +\epsilon v_2,v_1\}
}
$
\\
&&
${\scriptstyle
\{v_2+\epsilon v_1,v_3\},~\{v_3+\epsilon v_1, a v_3+ v_2\},~\{v_4+a v_1, v_3\},~\{v_4+a v_1, v_2\},
}
$\\
&&
${\scriptstyle
\{v_3,v_2,v_1\},~\{v_3,v_4,v_1\},~\{v_2,v_4,v_1\},~\{v_3,v_2,v_4+a v_1\}.
}
$\\
\hline

$k>\frac12,~k\neq 1$
&
${\scriptstyle
v_1=\frac{\partial}{\partial t},~
v_2=x\frac{\partial}{\partial u},~
v_3=\frac{\partial}{\partial u},
}
$
&
${\scriptstyle
\{v_3\},~ \{v_1\},~\{v_2 +a v_3\},~\{v_2+\epsilon v_1\},~\{v_3+\epsilon v_1\},~ \{v_4+a v_1\},
}
$\\
&
$
{\scriptstyle

v_4=x\frac{\partial}{\partial x}+(1-k)u\frac{\partial}{\partial u}.
}
$
&${\scriptstyle
\{v_2 +\epsilon v_3+a v_1\},~\{v_2,v_3\},~\{v_2,v_1\},~\{v_3,v_1\},~\{v_4,v_1\},~\{v_2 +\epsilon v_3,v_1\}
}
$
\\
&&
${\scriptstyle
\{v_3+\epsilon v_1,v_2\},~\{v_2+\epsilon v_1, a v_2+ v_3\},~\{v_4+a v_1, v_2\},~\{v_4+a v_1, v_3\},
}
$\\
&&
${\scriptstyle
\{v_2,v_3,v_1\},~\{v_2,v_4,v_1\},~\{v_3,v_4,v_1\},~\{v_2,v_3,v_4+a v_1\}.
}
$\\
\hline

\end{tabular}
\end{center}
\vspace{5pt}
\caption{The optimal system of subalgebras of the algebra $L$ (\ref{sysalg}) with $a\in \RR,$ $\epsilon=\pm1,~~0\leq\varphi\leq\Pi $ }
\end{table}

\section{Acknowledgments}
The author is grateful to Ljudmila A. Bordag for the posing of this problem and Nail H. Ibragimov for the interesting and fruitful discussions.   
\newpage

\end{document}